\documentclass[]{spie}  %>>> use for US letter paper
\usepackage{amsmath,amsfonts,amssymb}
\usepackage{graphicx}
\usepackage[colorlinks=true, allcolors=blue]{hyperref}
\usepackage{xcolor}

\title{Tearing and related field distortions in deep-depletion CCDs}

\author[a]{Claire~Juramy}
\author[a]{Pierre~Antilogus}
\author[b]{Laurent~Le~Guillou}
\author[a]{Eduardo~Sepulveda}
\affil[a]{LPNHE, (CNRS/IN2P3, Sorbonne Universit\'e, Paris Diderot), 
  Laboratoire de Physique Nucl\'eaire et de Hautes \'Energies,
  F-75005, Paris, France}
\affil[b]{Sorbonne Universit\'e, Paris Diderot, CNRS, IN2P3,
  Laboratoire de Physique Nucl\'eaire et de Hautes \'Energies, 
  F-75005, Paris, France}

\authorinfo{Further author information: (Send correspondence to C.J.)\\
C.J.: E-mail: juramy@lpnhe.in2p3.fr, Tel: +33 1 44 27 45 80, \url{https://orcid.org/0000-0002-3145-9258} \\  
P.A.: E-mail: p.antilogus@in2p3.fr \url{https://orcid.org/0000-0002-0389-5706} \\ 
L.L.G: E-mail: llg@lpnhe.in2p3.fr \url{https://orcid.org/0000-0001-7178-8868}\\
E.S.: E-mail: edo@lpnhe.in2p3.fr
}

\begin{document} 

\maketitle

\begin{abstract}
Tearing patterns affecting flat field frames in CCDs are a visually striking obstacle to performing Pixel Response Non-Uniformity corrections. These patterns can be explained by lateral field distortions, caused by the non-uniform distribution of holes in the channel stops between sensor columns. Over the course of LSST camera development, a number of practical fixes have been suggested to get rid of tearing. But applying these fixes to our 16-channel Teledyne-e2v sensors leaves at best a distortion pattern at the vertical edges of every segment. 

Our working hypothesis is that the origin of the tearing is the parallel clocking itself, which moves the holes that are present in the channel stops regions. The efficiency of these transfers depends strongly on the details of the clocking operations, resulting in the observed variety of distortion patterns.
Removal of most of the distortion patterns can therefore be achieved by executing a purge operation, which flattens back the hole distribution in the channel stops, immediately before acquiring a frame. A more effective solution is to switch all clocking operations to use a bipolar voltage set.
\end{abstract}


\section{INTRODUCTION}
\label{sec:intro}  % \label{} allows reference to this section
The Large Synoptic Survey Telescope (LSST) is building a 3.2 gigapixel camera for a wide-field survey in the optical and near-infrared range\cite{lsstdesign}. The survey design requires the 189 4kx4k pixels, 100\,$\mu$m thick CCD imaging devices to be fully read out in 2s\cite{lsstraft}. To achieve this, each device is segmented into 16 channels, with top and bottom serial registers divided into 8 segments. Constraints on sensor procurement have resulted into a mixed focal plane with sensors from two vendors, the CCD250 from Teledyne-e2v (T-e2v) and the STA-3800C from Imaging Technology Laboratory (ITL), resulting in some significant operational differences. One difference is that the T-e2v sensor specifications use a unipolar voltage set, with only positive voltages for clocks and biases\cite{e2v250}. The only exception is of course the Back-Substrate (BSS) bias voltage, which is applied to the back of the sensor to deplete it fully. Conversely, the ITL sensors are operated with a bipolar voltage set\cite{snyder18}, with negative lower clock rails and positive upper clock rails. 

When integrated with LSST electronics and operated in baseline conditions, all tested T-e2v sensors have produced flat field frames that are affected by tearing patterns, while none of the ITL sensors had this issue. There is some diversity in the observed patterns, as will be shown in the next section. But in all cases, this tearing creates an apparent local Pixel Response Non-Uniformity (PRNU), which compromises any future flat-fielding correction. As we will discuss, the tearing patterns are a symptom of lateral distortions of the drift field near the electron collection regions. Another consequence of this underlying issue is that we can expect strong astrometric distortions in the affected regions once the camera will be on the sky. 

Understandably, discussions of the tearing issue have usually focused on finding the right operational recipe to remove it, and are most often kept out of publications. One consequence of this approach is that the segment-edge distortion, which is related to the tearing, has been overlooked until now. Here we will first spend some time studying the field distortions, then focus on the exact conditions in which they are generated. With that knowledge, we will finally be able to propose two operational solutions to avoid these field distortions.

\section{OBSERVING THE TEARING PATTERNS}
\label{sec:obs}

\subsection{Experimental Setup}
The variety of tearing patterns, and their dependency on clock voltages and clock sequences, has been observed and tested during the qualification of the LSST science rafts, both at Brookhaven National Laboratory (BNL) and at SLAC National Accelerator Laboratory, by ourselves and by our LSST colleagues.
The more detailed investigations presented here have been performed on the CCD test bench at LPNHE, which has been described in Ref.\cite{astier19}. A single LSST-like T-e2v CCD250 is mounted inside a cryostat and cooled to -100\,C, while the LSST electronics board (REB) sits outside. A monochromator feeds into an optical fiber, which then feeds into an integrating sphere through a shutter, to provide flat field illumination to the sensor. Unless otherwise specified, flat field frames in this work have been taken at 650\,nm, with roughly 50,000~photo-electrons per pixel (with gain and offset, this translates to around 90,000\,ADU in raw frames). This value provides a good statistical significance, while still being halfway from the strong drop in the Photon Transfer Curve\cite{Guyonnet15}.

\subsection{Diversity of Distortion Patterns}
The notion of `tearing' covers different types of patterns, but they are unified by a common explanation, which also applies to the segment-edge distortions. Here we see how each pattern can be explained by variations of the density of holes within the channel stops between sensor columns.

\subsubsection{Wide tearing pattern}
\label{sec:wide}
   \begin{figure} [bt]
   \begin{center}
   \includegraphics[height=8.2cm]{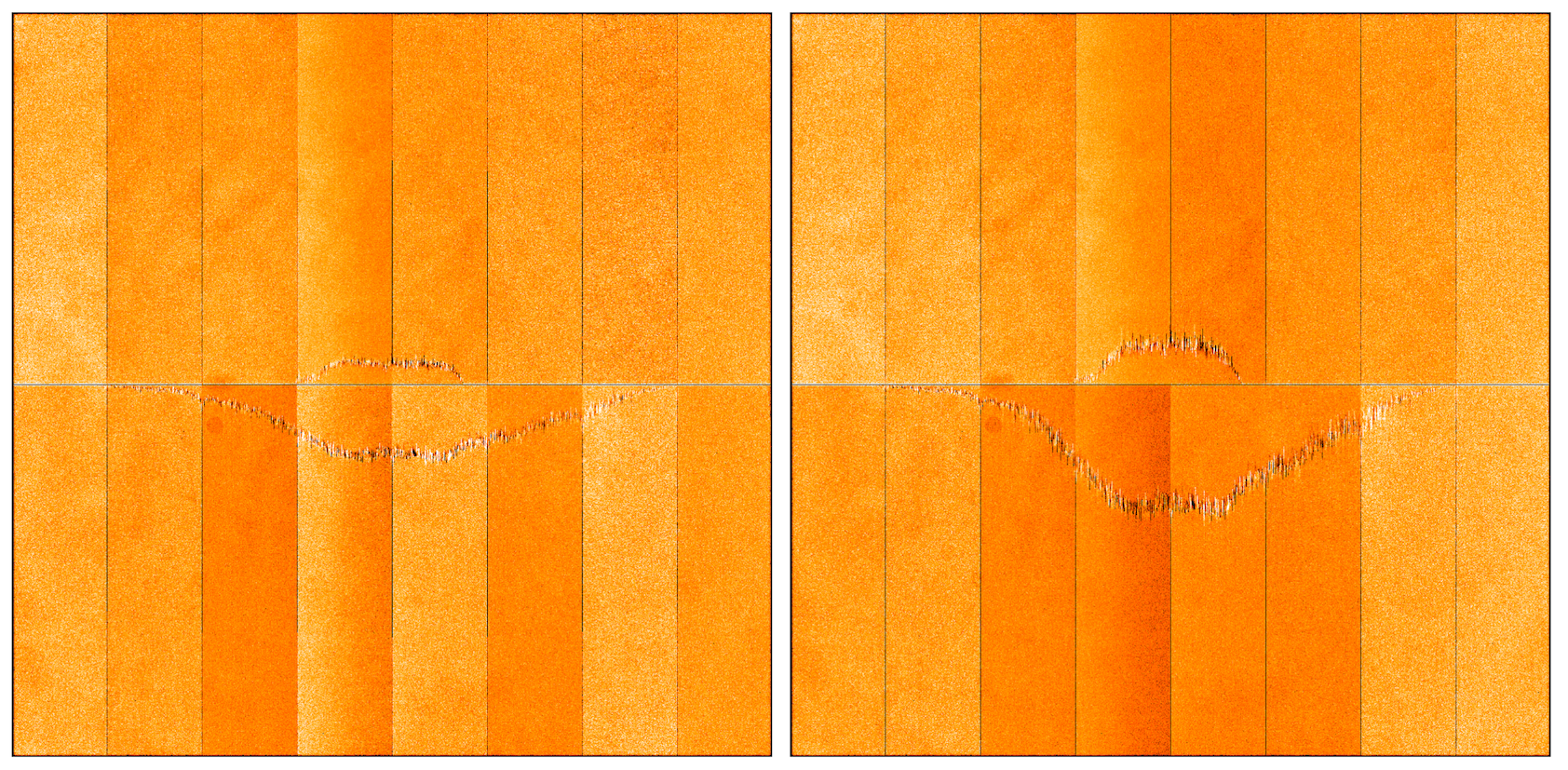}
   \end{center}
   \caption 
{ \label{fig:widepic} 
Two raw images, acquired successively with the 16-channel CCD, showing a wide tearing pattern over a flat field frame and its evolution.}
   \end{figure} 

\begin{figure} [tb]
   \begin{center}
   \includegraphics[height=6.8cm]{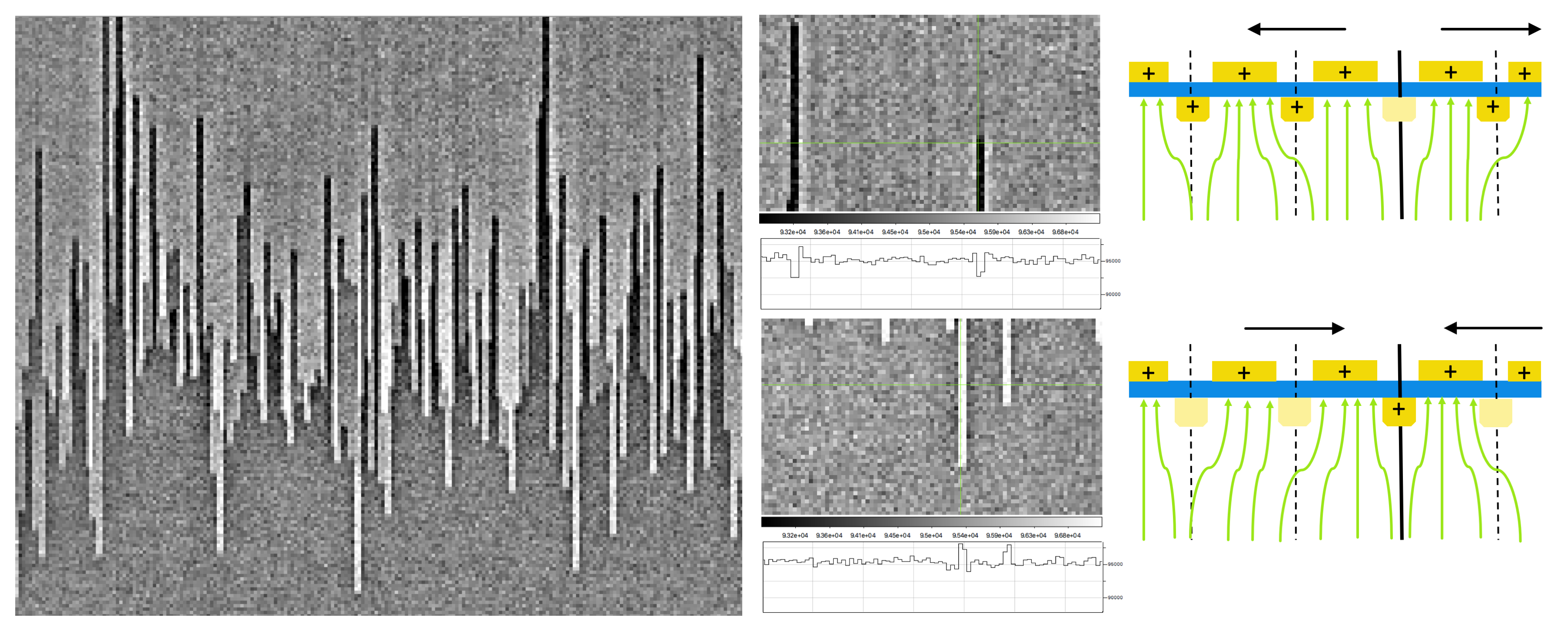}
   \end{center}
   \caption
{ \label{fig:skylinepic} 
Left: zoom on the previous `wide' tearing pattern, sometimes called a `skyline' pattern. The readout direction is toward the top of the frame. Center: profiles of isolated features, showing how darker (top) and lighter (bottom) columns always come in pairs, and how their neighboring columns are also affected, respectively with a light excess and a light deficit.  Right: rough sketch of electrical field lines in each situation, with the channel stop of interest marked with a solid black line. }
   \end{figure}

The most visually striking version of tearing spreads over hundreds of rows and columns on the frame, with a pattern width of tens of rows (Fig.~\ref{fig:widepic}). In LSST sensors, this wide tearing appears consistently when the upper voltage of the parallel clocks (PU, for Parallel Upper clock rail) is above 9.6\,V, with the lower voltage (PL) set at its default of 0\,V. This PU threshold is dependent on the BSS voltage, going down to 9.4\,V when BSS goes from -70\,V to -50\,V (Fig.~\ref{fig:pu2bss}).
To compound on the issue, when starting from a `blank slate' frame (as described in section~\ref{sec:study}), the pattern appears in the immediate next frame near the mid-line of the sensor (Fig.~\ref{fig:widepic}, left), and then moves up and down over consecutive frames, along the same directions as the parallel clocking, until it stabilizes. From the profile (Fig.~\ref{fig:skylinepic}), we measure a distortion amplitude of up to 7\,\% between neighboring columns. 
The peak-to-peak amplitude of the pattern remains constant from one frame to the next, and the spatial distribution of lighter and darker columns remain highly consistent, seeming to be only stretched along the column direction. 
The spatial distribution is also highly repeatable between sequences of frames, even with varying PU and BSS.

We also see in Fig.~\ref{fig:skylinepic} that all isolated features have in fact a width of two columns, which is the best visual evidence that the sources of the distortions are charges located in the channel stops. 
For instance, in the bottom picture, one channel stop holds a higher density of holes than its neighbors, creating a horizontal field near the collection regions that attracts the incoming photo-electrons to the two columns separated by this particular channel stop, and drawing them away from the next columns. 
This results in two bright columns, surrounded on both sides by columns that are slightly darker than average. 
At a given row, the feature disappears, signalling that the hole density in this particular channel stop has abruptly decreased, to the same level as its neighbors. 
Conversely, on the top side of this tearing pattern, most channel stops have a high density of holes. 
A channel stop with a lower density results in a horizontal field that draws photo-electrons away from the two columns separated by this channel stop, and into the next columns. 
Again, the feature disappears at the row at which the channel stop reaches finally the same hole density as the rest. 
In between, the bulk of the tearing pattern results from the combined fields created by several channel stops seeing their hole densities change abruptly at various rows in this region.

\subsubsection{Thin tearing pattern}
\begin{figure} [tb]
   \begin{center}
   \includegraphics[height=6cm]{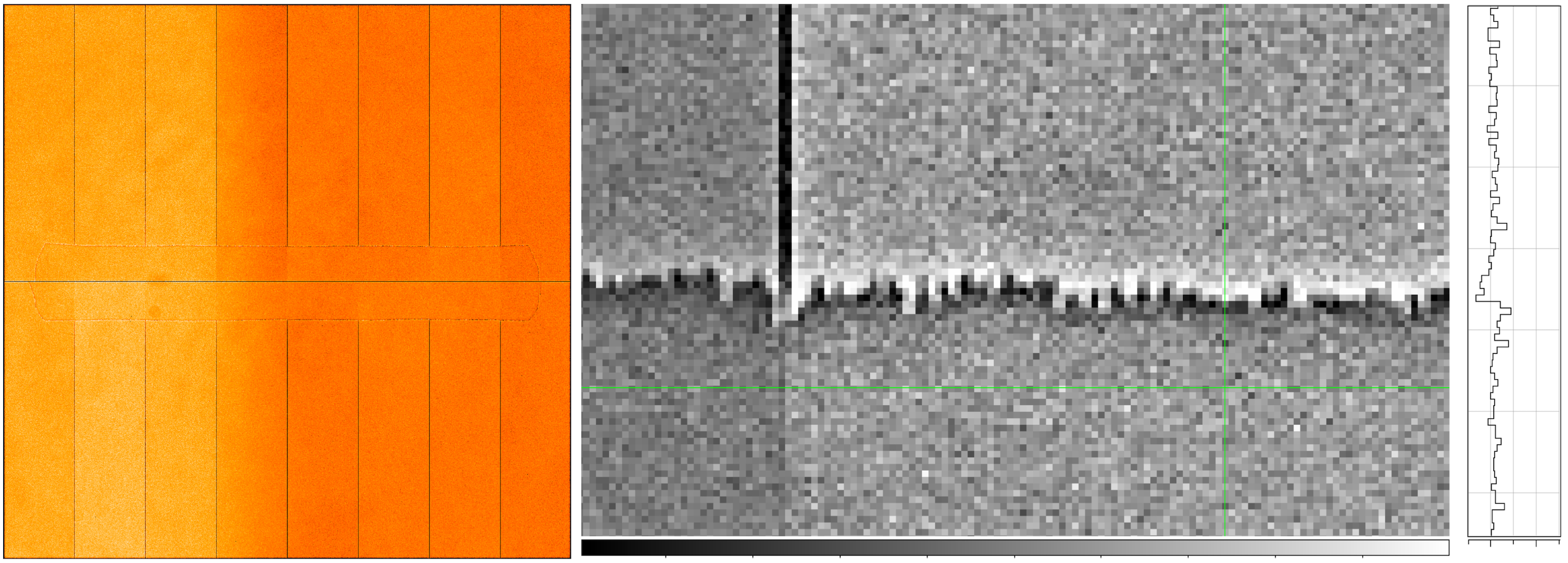}
   \end{center}
   \caption 
{ \label{fig:thin} 
Left: gain- and bias-corrected frame with a thin tearing pattern around the mid-line of the sensor. Right: vertical profile of the pattern (readout direction toward the top of the frame). Note also how the two dark columns marking the edge between the two readout segments go back to average levels below the tearing pattern.}
   \end{figure}

After unsuccessful attempts to remove the tearing pattern by changing the readout clock rates and clock timing, a first partial fix to the tearing issue was to lower the PU voltage. 
This resulted in a much thinner, nearly horizontal pattern, in a stable shape around the mid-line of the sensors (Fig.~\ref{fig:thin}). 
This pattern tapers off back to the mid-line before reaching the vertical sensor edges, which was also the case for the wider pattern. 

Again, we can explain this pattern in terms of hole density in the channel stops. 
There still has to be a steep transition of this density at some point along the rows, but now it happens in all channel stops within the same few rows. 
The end result is that, near this transition, there is a mostly vertical field distortion, so that incoming photo-electrons are attracted towards the rows with the higher density of holes in the channel stops. 
The profile of the transition shows that these rows with higher hole densities are found towards the top of the frame, when the parallel transfer for the CCD readout is done in the upwards direction, as was also the case for the wider pattern. 
The pattern is symmetrical in the lower half of the sensor. 

\subsubsection{Segment edge pattern}
\label{sec:rabbit}
\begin{figure} [tb]
   \begin{center}
   \includegraphics[height=5.5cm]{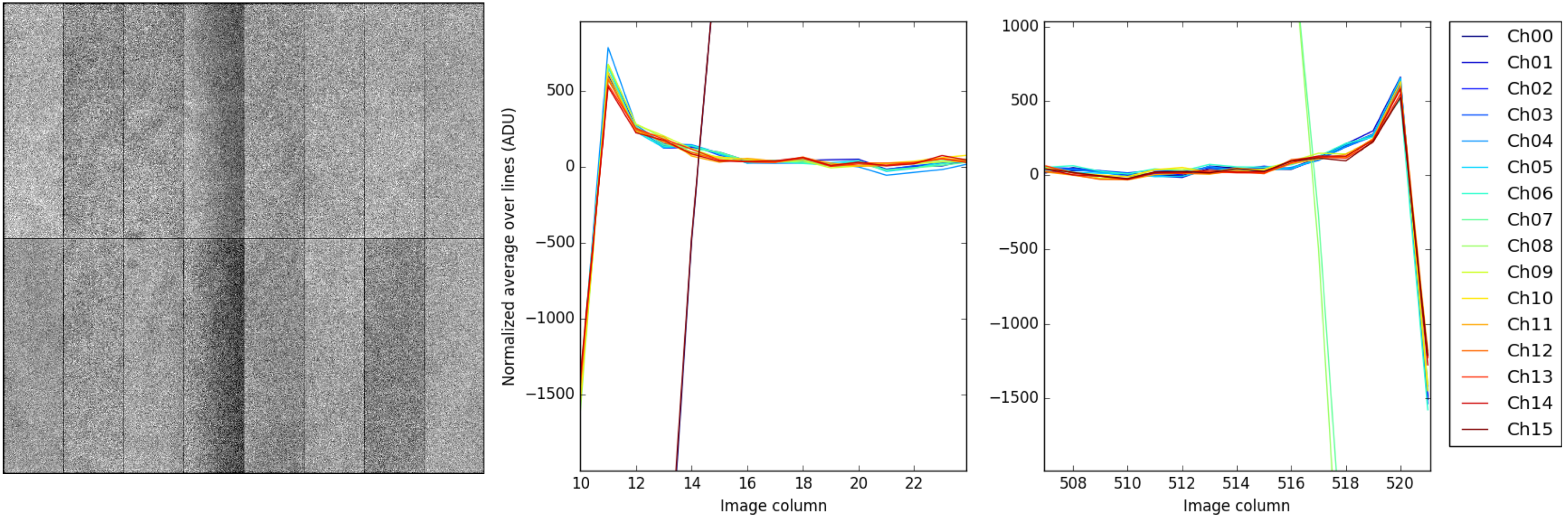}
   \end{center}
   \caption 
{ \label{fig:rabbitprofile} 
Left: raw flat field frame with no tearing pattern, showing dark columns at each segment edge. Center and right: profiles of the flat field at the columns near the edges of sensor segment, after averaging over rows and subtracting the flux away from the edge. This shows that the pattern we call the `rabbit ears' affect all 16 channels, symmetrically at the beginning (column 10) and the end (column 521) of the segments. The only exceptions show the edge roll-off at the vertical edges of the sensor.}
   \end{figure}

The next attempt to remove the tearing patterns involved modifying the timing of the fast clear sequence that is run before acquiring each frame. 
In this sequence, 2,010 parallel transfers are executed in succession to fully empty the frame, while all three serial clocks and the reset gate clock are held high to flush out any charges that arrive in the serial register. 
In this 4-phase sensor, these transfers are made up of four steps, with two phases moving in the opposite direction at each step. Each step originally lasted $6\mu s$. 
Following a suggestion from T-e2v, we lowered that duration to $3\mu s$ per step. 
This indeed removed the thin tearing pattern (Fig.~\ref{fig:rabbitprofile}), but at the cost of running the parallel clocks faster than the LSST electronics were designed for. 
Since the electronics could not deliver the required current, it is likely that the clock phases never actually reached the upper and/or lower voltage during those faster parallel transfers.

Besides, there is a remaining issue in this operating mode, which is localized at the vertical edges of each readout segment. 
All segment edges are consistently affected with a pattern we have called `rabbit ears' (see profiles in Fig.~\ref{fig:rabbitprofile}). 
This pattern has a peak-to-peak amplitude of the order of 3\,\%. 
It extends at least 5 pixels into the sensor, and is symmetrical across the segment edge. 
The symmetry excludes possible effects from the readout, like jumps in the electronic response, or some variant of serial Charge Transfer Inefficiency (CTI). 
This effect might be considered acceptable in most cases, but in our highly segmented 16-channel CCDs, those lateral field distortions would affect 2\,\% of the sensor collection areas. 

The `rabbit ears' pattern looks extremely similar to the dark columns of the wide tearing pattern. 
In addition, the `rabbit ears' are also found on all the frames that show tearing, but not uniformly over all rows: they stop at the tearing edge, as seen most clearly on Fig.~\ref{fig:thin}, and on the middle segments of Fig.~\ref{fig:widepic} (they are present on the other segments, but are smoothed out at that image scale). 
This also excludes that the rabbit ears are some intrinsic property of the pixels around the edges, that might have been an artefact of the sensor manufacturing process.

The `rabbit ears' can be explained similarly to the tearing effect, with the channel stop at the edge between segments having a lower density of holes than its neighbors. 
There is a possible origin in the sensor layout for the difference between this particular channel stop and the rest. 
All other channel stop implants end at the junction to the serial register, where their role is taken over by the S3 serial clock. 
On the other hand, the implant for the channel stop at the segment edge is prolonged beyond the serial register, being the boundary between the serial registers of the two segments (see T-e2v documentation\cite{e2v250}, p.7).

\subsection{Frames in Bipolar Mode}
\label{sec:bipolar}
Experimentally, we found two ways to suppress completely the tearing and the segment edge patterns. 
The first is to shift down all CCD voltages to use a bipolar voltage set, with $PL<-6V$. 
Then the first and last columns have a flux deficit reduced by a factor of ten, to roughly 0.25\,\%, and the neighboring columns are back to the average flux, which makes it quite possible that this leftover pattern is indeed an artefact of the sensor manufacturing process.

However, when the bipolar voltage set was tried on a 9-CCD LSST raft, this mode of operation failed to remove a wide tearing pattern in one particular sensor, which happened to exhibit a strong bulk defect. 
The second method, which involves pulling all four phases of the parallel clocks together down to around -7\,V for a duration of 2\,ms\footnote{Disclaimer: T-e2v has expressed a number of reservations about executing this operation on CCD250. We do not advise to attempt it on the sensors they produce without consulting them first.} did work on all nine sensors. 
This operation, executed immediately before acquiring a frame, is an attempt to `purge' holes from the channel stops. 

Two similar operations are commonly implemented in the case of p-channel CCDs\cite{lbnl, rahmer12}, the `erase' procedure to eliminate residual images, and the `e-purge' procedure to purge charges from the channel stops, which create fixed patterns. 
The `purge' operation we use here would be the equivalent operation to an `erase' procedure, taking into account the inverted polarities, but we do not ramp BSS at the same time. 
As for n-channel CCDs, in the documentation for CCD290-99\cite{e2v290}, T-e2v describes a similar operation to our `purge', either lowering the parallel clocks to -10V, or increasing the Front Substrate voltage to +10V, and recommends this operation to remove the tearing (note 19, p.17). However, it should be noted that, unlike CCD250, CCD290-99 does not have a BSS applied.

In our case, it is not certain that we actually reach inversion at this voltage, or really remove the holes. This may be why, unlike what is reported for p-channel CCDs, it is necessary to repeat the operation before each frame. 
But, at minimum, it seems sufficiently low to draw the holes away from the channel stops, and to redistribute them uniformly across the sensor.

\subsection{Field Distortions from Channel Stops}
Since the tearing is affected by the parallel clock voltage and by the timing of the transfers, one alternative explanation we considered was an issue with the parallel charge transfer, in particular at the point of transfer to the serial register. 
But the fact that the excess of charges comes first in the readout direction allows us to discard this possibility. 
Similarly, the disappearance of the `rabbit ears' pattern at the tearing edge allows us to rule out a transfer issue.

Observing the various patterns, and their disappearance, we have already seen that they can be explained by considering non-uniform hole distributions in the channel stops. 
This explanation has already been presented elsewhere, for instance in the same note in the CCD290-99 documentation\cite{e2v290}. 
This effect can be considered the counterpart of the brighter-fatter effect: in the brighter-fatter effect, electrons are collected in the potential well of the collection region, and their accumulation modifies slightly the local electrostatic field, decreasing the effective pixel size of the pixels with the higher counts of electrons. 
Conversely, the implants in the channel stops create regions that are repulsive to electrons, and holes can accumulate in the (inverted) potential well near the sensor front surface. This accumulation of holes then modifies slightly the local electrostatic field, increasing the effective pixel size of the pixels on either side of the channel stop with the highest count of holes. 

The amplitude of the distortions is of the same order of magnitude than that of the brighter-fatter effect. 
To compare, the highest correlation coefficient in flat fields due to brighter-fatter is $C_{0,1}$, the correlation with the next pixel along the vertical direction. 
Its value is 0.015 at a flux level of 50,000\,e-, for the same sensor in the same conditions\cite{Guyonnet15}. 
But one key property of the brighter-fatter effect is that this correlation increases as the flux comes in. 
To better understand the origin of the tearing, and of the related segment-edge distortions, we will now do a similar study on their amplitude.

\section{STUDY OF THE FIELD DISTORTIONS}
\label{sec:study}
First, we will test how the hypothesis that the patterns are due to field distortions holds up experimentally. Then, we will explore the dependency of those field distortions on the exact operating conditions of the sensor.
When performing these tests, we distinguished as much as possible between the operations that generate the field distortions, and the observation of the result, that is the tearing pattern in a flat field frame, and the measurement of its amplitude. Our metric is the average flux in the first and last columns of each segment, normalized to the flux at 10+ pixels into the segment. 
Since tearing can evolve over repeated frames, we need to start from a `blank slate' state, so that the patterns that appear can be reliably attributed to the operations done afterwards. 
As we discussed in section~\ref{sec:bipolar}, one way to produce such blank slate state is to execute the purge operation at -7\,V.

\subsection{Characterizing the Field Distortions}
In this section, to generate a `rabbit ears' pattern in a repeatable way, we purge, and acquire a 50,000\,e- flat field at 650\,nm.
Then we observe the field distortions in the next frame, for various exposure times (in a randomized order).
The normalized amplitude of the first and last column in this second frame increases very slightly, which means that the amplitude of the `rabbit ears' decreases, when the flux in the second frame increases. 
This change can be explained by the brighter-fatter effect, which increases the correlation between neighboring pixels at higher fluxes. 
In this case, the net effect of this correlation is that a fraction of the flux that should arrive in the brighter column is instead pushed to its darker neighbor at the edge, due to the repulsive effect of the charges already accumulated. 
This fraction is proportional to the difference in flux between these two columns. 

Numerically, the average difference over all channels is 3.1\,\% of the flux, and this difference decreases on average by 0.14\,\% ($\pm 0.03$\,\% over all channels) between the frames at 35,000 and 120,000\,e-.
Since the patterns are columns, we should have a good estimate of the correlation by adding up the contributions from the pixel in the next column ($C_{1,0}$), as well as its immediate top and bottom neighbors ($2\times C_{1,1}$) from Ref.~\citenum{Guyonnet15}. 
This total correlation increases roughly from 0.6\,\% to 3\,\% between the two fluxes. When applied to the difference between the two columns, this gives: $3.1\% \times (3\% - 0.6\%) = 0.074\%$ as the fraction of the flux subtracted from the bright column and added to the dark column at the higher flux. The overall effect is a decrease in the difference of 0.148\,\%, fairly close to our measurement.

So, aside from this brighter-fatter correction, the amplitude of the tearing pattern is completely independent of the flux measured in the frame that exhibits the pattern. 
This is compatible with the idea of a pre-existing field distortion, that is not affected by incoming light.

Another tool to validate the field distortion model is to change the BSS voltage. 
When its absolute value is lower, the drift field from the back of the sensor to its front is weaker, so the lateral field distortions have more effect on the final distribution of electrons in the pixels. 
This is illustrated when comparing patterns at -70\,V and -50\,V in Fig.~\ref{fig:pu2bss}.

\subsection{Generating Distortions with Parallel Clocking}
\label{sec:genclocks}

   \begin{figure} [tb]
   \begin{center}
   \begin{tabular}{c} %% tabular useful for creating an array of images 
   \includegraphics[height=7cm]{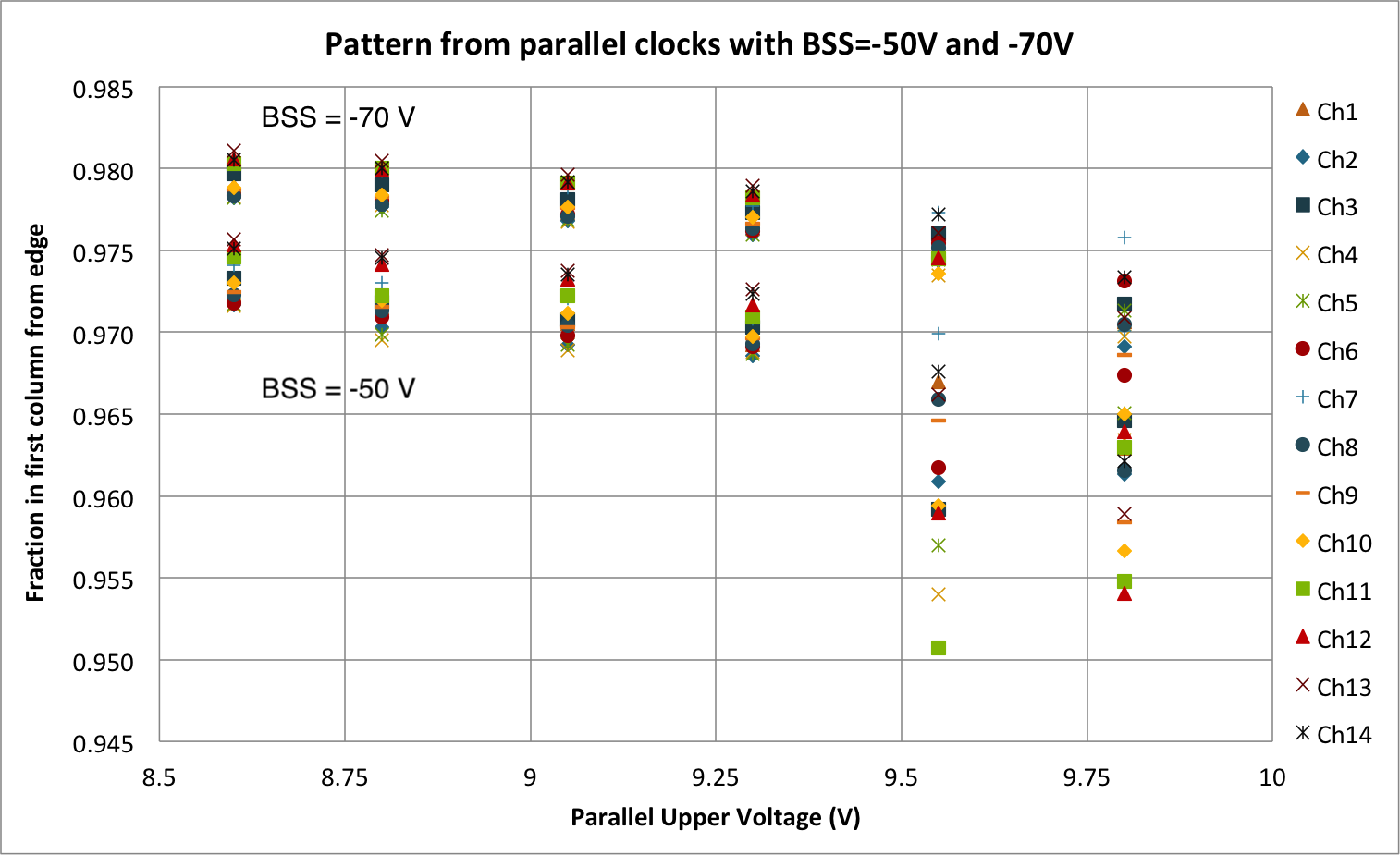}
   \end{tabular}
   \end{center}
   \caption 
{ \label{fig:pu2bss} 
Amplitude  of the `rabbit ears' pattern in the last column of each segment, when varying the PU voltage in the previous frame. In the two BSS sets, the different BSS voltages were applied to both the previous frame and the observed frame. At higher PU, the amplitude measurements become highly dispersed due to contamination by the wide tearing pattern.}
   \end{figure} 

Here we tried various sequences of exposure and readout modes, to pinpoint how the field distortions emerge. 
After a purge at -7\,V, if the frame is exposed and read out right away, there are no distortions, independently of the voltage set that was used.
If a clear was performed with the unipolar voltage set (any PU), the `rabbit ears' appear. 
Then, the state of the second frame depends on the voltage set used to read out the first frame, and on whether a clear was performed before acquiring the second frame, but is completely independent from the voltage set used during the acquisition of the second frame: 
\begin{itemize}
    \item First frame read out with bipolar set, no clear: no distortions.
    \item First frame read out with the unipolar set at low PU, no clear (or extra fast clear with the same, as in \ref{sec:rabbit}): `rabbit ears' appear.
    \item  First frame read out with the unipolar set at low PU, then clear with the same: thin tearing pattern.
    \item First frame read out with unipolar set at high PU, no clear or clear at same PU: wide tearing pattern. 
\end{itemize}
In all cases, the tearing patterns appear when clocking operations were executed with the unipolar sets before exposing the frame. The voltage set used during exposure of either frame has no impact on the patterns. 
From longer sequences of frames, we have also seen that executing a read or a clear with the bipolar set is sufficient to wipe out the pattern for the frame that comes immediately after.

We also performed quantitative tests on the effect of the PU voltage, by repeating the following sequence: purge at -7\,V, acquire a flat field frame with the desired PU, then shift back to the baseline PU (9.3\,V) to acquire the next frame, with no clear before either frame.
As the PU voltage used to acquire the previous frame increases, the amplitude of the`rabbit ears' in the next frame increases (Fig.~\ref{fig:pu2bss}), until the large tearing pattern appears.

\subsection{Other Contributions to Field Distortions}
While the tearing patterns are clearly tied to the parallel clocking operations, we have found other contributing factors for the amplitude of the distortions, using the same test method. For instance, when increasing the integrated flux in the first frame, the normalized flux in the last columns of the second frame decreases linearly, which means that the amplitude of the pattern increases. Overall, between the fluxes of 0 and 120,000\,e-, the pattern increased by 0.2 to 0.3\,\% depending on the segment, a relative increase of 10\,\%. 

It might be tempting to attribute this increase to holes generated by the creation of electron-hole pairs from incoming photons, but it is more likely to be caused by some side effect from increasing the exposure time. 
To compare, just waiting in the dark after the purge seems to add to the edge distortion by up to 0.1\,\%. 
In both cases, the serial register clocks are run periodically during that time, and again right before reading out the frame. As we will see in Sec.~\ref{sec:unipolpurge}, the exact configuration of the serial clocks has some effect on the patterns.

\section{MINIMAL WORKING MODEL}
The goal here is to make a minimal model that accounts for our observations. 
Our starting point is that in unipolar mode, there are holes confined to each channel stop, preferentially underneath the parallel clock electrodes that are set to the PL voltage.
After a purge at -7\,V, when shifting back to the unipolar voltage set, the holes either come back, or they had never really left, and are trapped into the channel stops with a distribution that is uniform from pixel to pixel, or at least vary only on large scales. 

Then, we need a mechanism to move the holes to break this uniformity: this is the parallel clocking, which should be moving the holes in the same direction as the electrons. 
In the channel stops at the segment edges, the holes arriving at the level of the serial register can get out through the implant, unlike in other channel stops where they are blocked from going into the serial register. 
This is because the serial register is a buried channel (as illustrated in Figure~1.12(a) of Ref.~\citenum{janesick}), which potential facilitates by design the transfer of electrons from the imaging area to the serial register, and conversely repulses holes. 
This difference can explain the standalone `rabbit ears' pattern. It does not tell us whether the holes are in fact moving only in the segment edge channels, or if they can move in all channels, but are in effect backed up at the serial register. 
In this second case, we can suppose that once the number of holes in the next row reaches a certain level, they create a repulsive field strong enough to prevent any further transfers. 

For the transition from the `rabbit ears' to the tearing patterns, we need the  efficiency of the hole transfer to vary, so that only some variants of clocking can move the holes efficiently enough to empty the channel stops, starting from the mid-line, or at least enough to create a discontinuity in the hole distribution.
In the case of the wide tearing pattern, the increased efficiency comes from the higher PU voltage, which must be more efficient at pushing holes away. 
The evolution of this type of tearing in the first few frames is the sign that the edge of the hole density moves in the readout direction. The tearing pattern then stabilizes, either because it is blocked by the repulsive effect of the accumulated holes in the next rows, or because the efficiency of the hole transfer varies across the surface of the sensor. 

To explain that the thin tearing pattern appears only when a clear is performed (and not a readout), we have to suppose that our clock sequence used for clearing is more efficient at moving the holes. 
This is reasonable, because the parallel transfer sequence for the readout was modified specifically to transfer the electrons more efficiently, adding some overlap times with three phases high out of four, instead of having only two phases high at any time, as in the fast parallel transfer used for clearing. 
While this did improve electron transfer during readout, particularly at high flux, it would necessarily decrease the efficiency of the hole transfer. 
As an additional confirmation, we have managed to move the position of the thin tearing pattern between different rows, depending on the exact timing of the clearing sequence. 

There remains the question of why the tearing patterns are non-uniform across the sensor columns. On large scales, one possible explanation comes from the propagation of the parallel clocks across the surface of the sensor. 
We know that the clock waveforms are smoother near the middle of the sensor than in the corners, which tends to make them less efficient to transfer electrons. 
Again, by symmetry, this might mean that the hole transfer is then more efficient. 
Other explanations could be a large-scale non-uniformity in the hole distribution, or a difference in the capacity of channel stops either to hold holes, or to allow their movement. 

These other explanations apply also to small-scale differences in the hole distribution after clocking, as seen in the `skyline' pattern.
In that particular case, since the distribution of lighter and darker columns is stable from one sequence to the next (as mentioned in section~\ref{sec:wide}), the underlying explanation must be tied to the intrinsic properties of the sensor. 
One such property could be the variation of the local dopant density in the channel stops. 
Such variations on larger scales could be another way to account for the large-scale non-uniformities of the patterns.

   \begin{figure} [tb]
   \begin{center}
   \includegraphics[height=7.0cm]{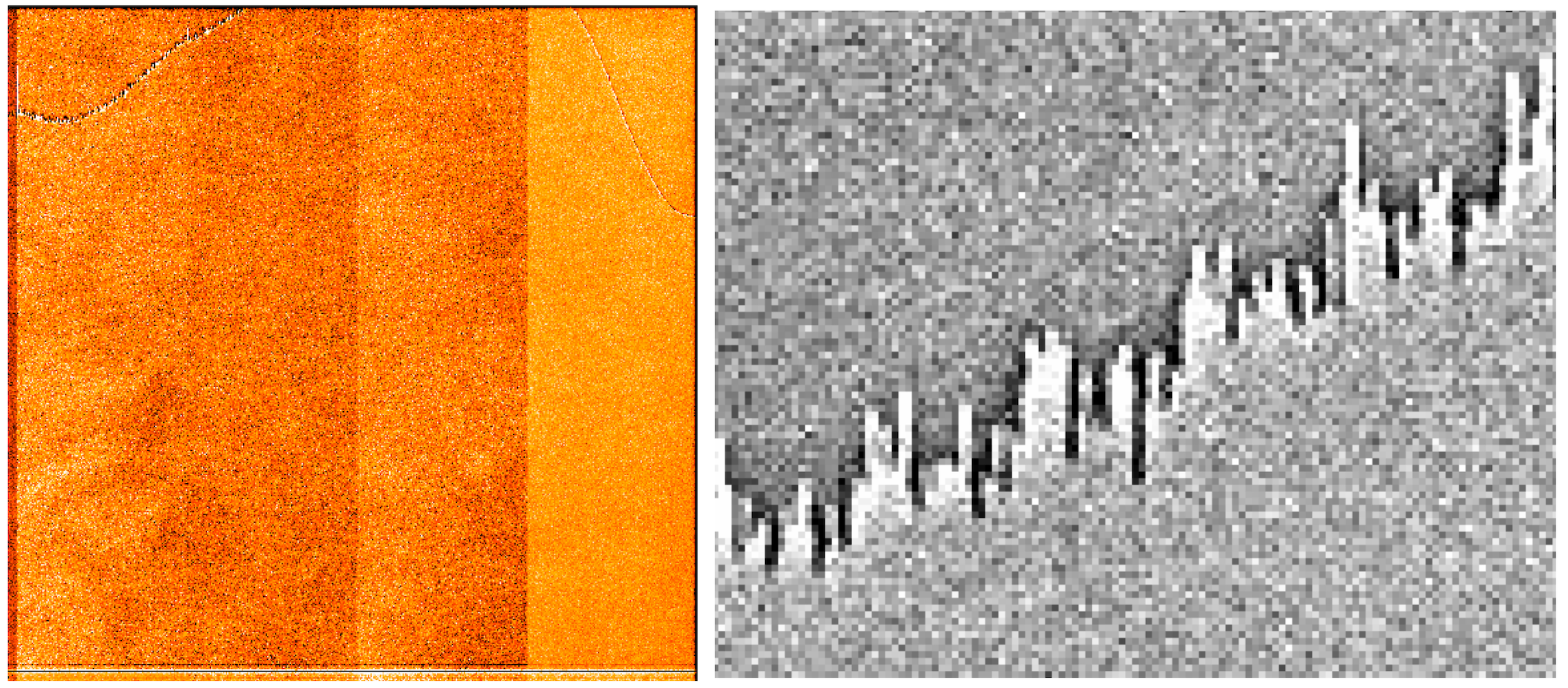}
   \end{center}
   \caption
{ \label{fig:reverse} 
Left: flat field (top right quarter of the sensor) obtained after a purge at -7\,V and reverse parallel clocking for 1,000 rows (half a frame), with PU voltage at 9.8\,V. Patterns appear both at the corners of the sensors, and in the middle segments near the sensor horizontal edges. Right: zoom showing the reversal of the tearing pattern, with dark double columns towards the bottom of the frame and bright double columns towards the top (readout direction is towards the top of the frame).}
   \end{figure} 

To validate some aspects of our models, we ran the parallel clocks in reverse, with the unipolar voltage set, right after a purge at -7\,V. 
In the first frame, a tearing pattern appears, in reverse to the usual patterns (Fig.~\ref{fig:reverse}): it is located close to the edges of sensor, instead of the mid-line, and the tearing transition indicates that the hole density is lower in the outer areas, instead of being lower around the mid-line. 
This means that the holes were present in the sensor as soon as we shifted back to the unipolar voltages, and that we were able to move the holes away from the serial registers with this reverse parallel clocking. 
In addition, we performed this reverse clocking both with high and low PU voltages. 
With the low PU, the reverse tearing pattern appeared only in the corners of the sensor. 
With the high PU, a wider reverse pattern appeared also in the middle segments, confirming that this higher voltage is more efficient to move the holes. 

\section{SOLUTIONS FOR REMOVING PATTERNS}

\subsection{Operating in Bipolar Mode}
As discussed in Section~\ref{sec:bipolar}, one way to prevent the tearing patterns from ever appearing is to operate the sensor with a bipolar voltage set. 
As seen in Section~\ref{sec:genclocks}, operating in bipolar mode only during the readout phase, or just executing a purge operation at -7\,V before acquiring a frame, has the same effect. 
Our preliminary studies indicate that operating the readout in bipolar mode would also suppress an abnormally high, flux-dependent serial CTI that appears in some channels of our sensors. 
We can also expect that it would slightly improve dark current. One other advantage is that it removes the constraint that $PU-PL<9.5V$ (with PL constrained at 0\,V in unipolar mode), which opens up the parameter space to optimize the CCD full well.

While our LSST raft electronics can deliver the bipolar voltage set, the operating parameters of the sensor are highly constrained. 
One such constraint is that by construction, the parallel clock phase P4 is also used as the reset of the second stage of the CCD output amplifier\cite{e2v250}. 
Finding the optimal bipolar voltage set for all T-e2v sensors on the LSST focal plane is a work that is not complete, and out of the scope of this article. 
Even switching briefly the voltages to apply a purge would create significant overhead relatively to the 2~s readout time of LSST, so there was a strong interest in finding a tearing fix that could be used at the baseline voltage set.

\subsection{Purging in Unipolar Mode}
\label{sec:unipolpurge}
In unipolar mode, we cannot really `purge' the holes, but our goal is to spread them out uniformly, or at least with no sharp transitions, in a repeatable manner. 
We have found experimentally that this can be achieved by pulling all parallel clocks down to PL, which is set at 0\,V. 
This applies a uniform potential along each channel stop, instead of having pockets of potential that depend on which parallel phases are at PU or PL.
This operation removes the tearing patterns, including at high PU, and most of the segment edge distortions, in the frame acquired immediately afterwards. 
We only have a faint `rabbit ears' pattern left near the serial register, which is limited to less than 200 rows, with an amplitude of at most 1.5\,\%. 

But this operation, equivalent in effect to a `purge', works only if certain conditions are met. 
First, it must be kept in this state for a long enough time, at least 3\,ms for our sensor. 
This time scale means either that the channel stops are not really conductive for holes, so their propagation takes a non-negligible time, or that the holes are trapped at the surface and need some time to be released. 
Then, since the parallel clocking is what generates the segment edge pattern, this must be the last operation before acquiring a frame, done after clearing. 
Also, since the later operations of readout and clearing will generate a non-uniform hole distribution again, this `purge' must be performed before each frame, ensuring repeatability.
Less obviously, we have found that the serial clocks S1 and S2 must be set high at the time of the purge to obtain the minimal residual pattern, while the state of S3 seems indifferent. 
It is possible that when these two clocks are high, they affect the potential of the segment edge channel stop at the point where it crosses the serial register, decreasing the probability for holes to drain away through that path. 

This unipolar purge-like operation has been tested on an LSST raft, with the exact same result of removing the tearing patterns, and leaving only a fraction of rabbit ears, on all nine T-e2v sensors. It will be incorporated into the baseline operations for the LSST readout. 

\section{CONCLUSIONS}
Tearing and the related `rabbit ears' pattern have been long-standing issues in the LSST rafts built with T-e2v sensors, with no adequate solution up to now. 
These patterns are created by lateral distortions of the electric field, that affect incoming photo-electrons drifting to the accumulation regions. 
These field distortions are caused by non-uniform hole distributions in the channel stop regions.

The parallel clocking of the frame during the readout and clearing operations is the main source for these non-uniformities, but there is evidence for additional effects at play. 
A purge-like operation with the usual unipolar voltage set clears the tearing and most - but not all - of the `rabbit ears', if executed correctly. 
Since this solution applies even at the higher parallel clock voltages, it opens up a new opportunity for sensor optimization, by removing one of the constraints on this voltage set. 
Moving to a bipolar voltage set would remove the issue entirely, and would have other beneficial effects, but this would require extensive study and validation.

Beyond the operation of the LSST sensors, one promising avenue to lift the uncertainties in our model of tearing would be through modeling of the sensor physics, and possibly through simulation to replicate hole movements. 
On the experimental side, comparison with the strategies used on other sensors to remove similar issues may yield new configurations to try, which might in turn improve our understanding of the underlying physics.

\acknowledgments % equivalent to \section*{ACKNOWLEDGMENTS}       
 
The authors thank their colleagues in the LSST project and in the Dark Energy Science Collaboration (DESC) for many productive discussions, and for their feedback on this study. In particular, we thank Craig Lage at UC Davis for sharing the results of his work on modeling the electrical properties of the LSST sensors.

% References
\bibliography{mybib} % bibliography data in mybib.bib
\bibliographystyle{spiebib} % makes bibtex use spiebib.bst

\end{document}